# Assessing usability of model driven development in industrial projects

Florian Fieber[1,3], Nikolaus Regnat[2], and Bernhard Rumpe[1]

[1] RWTH Aachen University, Department of Computer Science 3,
Ahornstraße 55, 52074 Aachen, Germany
rumpe@se-rwth.de

[2] Siemens AG, Corporate Technology CT SE 1,
Otto-Hahn-Ring 6, 81739 München, Germany
nikolaus.regnat@siemens.com

[3] qme Software GmbH
Gustav-Meyer-Allee 25, 13355 Berlin, Germany
florian.fieber@qme-software.de

**Abstract.** An integral use of the model driven development paradigm influences and changes an organization's software development division rather heavily. Such a paradigm reduces some tasks in complexity and costs, but also introduces new tasks and, if introduced seriously, has severe affects on activities and roles in the software development process. As the model becomes the most important development artifact, there are new challenges to the development team, e. g. assessing the model's quality, model partitioning and configuration management for distributed teams, setup of build management, tool chaining and tracing of information through the various artifacts. Organizations coping with model driven development need to successfully introduce new tools and new ways of thinking, they are challenged in adopting their processes and training their staff. This paper presents an ongoing research project on the assessment of the usability of modeling and model driven development at a global industrial organization with its headquarters in Germany. The matter of interest is the analysis of the usability of modeling (especially with the UML) and model driven development by accomplishing an empirical, quantitative survey.

## 1 Introduction

The Unified Modeling Language (UML) is a loosely coupled set of object oriented modeling notations, standardized by the OMG [20,21]. It has become a de-facto standard and been successfully adopted by industry for the specification and documentation of object oriented systems [8,9]. The UML is a modeling notation but does not provide a method on how to model respectively adopt modeling in the software development process. However, most software process models consider modeling activities (e. g. design of the software architecture), roles (e. g. architect/designer) and outcomes (e. g. software architecture/design), but in a variety of ways. Albeit there is an intense use of software models and large parts of the system are modeled, the process is still code centric as the code has almost always to be developed manually.

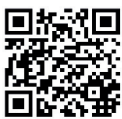

With the advent of the model driven development paradigm (e. g. OMG's Model-Driven Architecture approach [18,19]) the model becomes the central development artifact and the process becomes model driven as code and documentation are (ideally) being generated from the model. While in code centric processes the model is a sole means for documentation and specification, in model driven development approaches the model becomes a "real" development artifact and modeling activities have to be considered in the development process, e. g. model reviews in quality assurance activities. However, the MDA approach does itself not provide very much methodology for the adoption of the model driven development paradigm in the software development process. Companies applying the model driven development paradigm are challenged to adopt their existing processes in respect to these new and changed activities.

Those challenges are various, starting with extended technological issues, such as a necessity for homogeneous, much more elaborated development environments, a wider variety of development artifacts that want to be managed, stored, versioned, generated or have other kinds of dependencies. This leads to a larger set of "roles" in the development process and it has to be clarified, whether and how these roles are assigned to people (including individual or common ownership of artifacts). Siemens is a company with one of the largest software development companies and we can expect to learn a variety of solutions as well as current challenges that occur when adopting the model driven development paradigm.

The remainder of the paper is organized as follows. Section 2 presents the status quo of modeling and model driven development at Siemens. Section 3 presents our research project, setup of interviews and first findings. Section 4 summarizes and concludes the paper.

## 2   Software Development at Siemens

Siemens is one of the world's largest software companies but typically is not recognized as one, as most of the software is part of embedded systems. Nevertheless, Siemens has more than 20.000 software engineers worldwide and around 60% of the Siemens business is based on software.

### 2.1   Organizational Structure

In 2007 Siemens changed its internal organization and now has three sectors: Industry, Energy and Healthcare. Each sector has a broad range of products and/or provides solutions where software is an integral part – from industry automation systems to power distribution and medical products. Some software development departments within these sectors are developing software since decades but increasing complexity and introduction of globally distributed teams create new challenges that have to be solved. Improving the software development processes and searching for solutions to these new challenges has been an important topic over the past decade. In 1995 Siemens established a best practice sharing forum, the Siemens Software Initiative, which addresses strategic software topics and shares technical best practices regarding processes and architecture.

### 2.2 Model Driven Development at Siemens

During the last five years the model driven development paradigm began to influence the software developing departments within the Siemens sectors. Although languages like the UML are available for more than a decade, it took years until the industry picked them up. Currently the languages and tools are finally mature enough to be introduced in an industry environment but the basic principles on how to handle models or introduce model based approaches in industry scope projects are not well known or documented. As a result, at Siemens there's currently no common approach regarding model driven development – one reason may be the huge diversity of produced products and provided solutions within Siemens. Currently each software development department has to define its own way of working with models (e. g. [11]). As most model driven approaches however have similarities and, even more important, share common challenges, a goal at Siemens and especially at the Siemens Software Initiative is to analyze and consolidate the topic of model driven development. Building up knowledge on the basic principles when introducing a model based approach to a software department and sharing this knowledge is a major goal during the next years. Understanding and documenting the key challenges and success factors will be the first step to reach this goal.

### 2.3 Key Challenges and Success Factors

When introducing a model driven development approach a number of key challenges have to be solved. Choosing the appropriate modeling language is the first and most important step. Using e. g. the UML out of the box is rarely possible as most software developers working on embedded systems are electrical engineers and are seldom familiar with the UML. In several projects it turned out that customization of the UML ensures that the language is accepted by the developers – this can be done by restricting the elements and diagrams to be used and by creating domain specific profiles.

As the UML provides no method it is crucial to define one during introduction of an MDSE approach. Typically there are already existing software development methods and processes within the Siemens departments – therefore the challenge is to integrate a model-based approach with minimal impact. Especially the interface to other departments is critical – it rarely happens that all departments introduce such radical changes simultaneously. Using customized document generators is one possibility to ease the transition between traditional and model based development approaches.

Another important step is to decide on how to organize (i. e. partition and structure) models. Once a tool has been selected that supports the defined approach and also takes the existing tool infrastructure into account, the organization of models is critical. Distributed world-wide development and large teams are common within Siemens software departments and without a proper organization for models an MDSE approach is bound to fail.

The given examples are just a few of the typical challenges one has to deal with when introducing model driven software development at industry scope.

## 3 Empirical Study

There are a number of supposed key benefits of model driven software development which are constantly reported, e. g. increasing productivity, improving quality, standardization and formalization in [17]. However, there are only few investigations on these statements in large projects and almost no valid empirical data available. Besides, only some approaches discussing the integration with light- or heavy-weight process models exist [3,4]. Furthermore, as described above, modeling as well as model driven development activities are executed but rarely handled in formal software process models in industry.

### 3.1 Related Work

There is not very much work that relates to empirical investigations of modeling or model driven development in a larger scale, i.e. beyond reports on parts of individual projects. Most work relate to modeling conventions or the quality aspects of certain UML diagram types.

Asadi and Ramsin [16] provide a review of several model driven methodologies and present a criteria-based evaluation and a framework for assessing, comparing, selecting, and adapting model driven methodologies. They compared and evaluated different MDA-based methodologies and concluded:
- The methodologies are not mature enough, especially in regard to supporting standard software engineering activities. Besides, definitions of the methodologies are not complete.
- Most of the methodologies do not offer any guidelines on the usage of MDA tools in coherence with the methodology. All tool-related issues are left to the tool vendors.
- Most methodologies do hardly provide other important MDA features like support for extension of rules, round-trip engineering, model synchronization. and model verification/validation.

Mohagheghi and Dehlen [17] reviewed 25 empirical studies by evaluating reasons for and effects on applying the model driven development paradigm in industrial projects. Among others, they present the following findings:

- Increasing productivity (and shortening development time) and improving quality may be regarded as the ultimate reasons for applying the model driven development paradigm.
- Most known processes are not tailored for the model driven development paradigm. Besides, the paradigm does provide any support for the software development process or the design methodology. It may be unrealistic to use a pure model driven process as "software engineering methods are not fitted to use models as main artifacts, i. e. activities such as analysis and evaluation is still largely done at the code level" and "software engineering environments are not mature enough."
- Some projects suffer from productivity loss due to immature tools and high start up costs, and that modeling can be at least as complex as programming. Also, reasons for not adopting the paradigm are high initial investment and unsure benefits.
- However, "most papers evaluate models as useful for improving understandability and communication among stakeholders".

In any case they identify a demand for more empirical studies and detailed data, as only few reports are on larger projects. Return-On-Investment (ROI) aspects should be evaluated in future as well..

Lange [12] presents a comprehensive set of 21 papers on modeling quality in his dissertation. The work covers papers on model quality attributes, modeling conventions, metrics and quality assessment.

Wang and Brooks [13] describe three empirical studies on conceptual modeling and the modeling process. The studies analyze the efforts of modeling and effects on the modeling process.

Dzidek, Arisholm and Briand [14] accomplish a controlled experiment that investigates the costs and benefits of using UML documentation in an industrial project.

Anda and Hansen [15] describe a case study on the application of UML in legacy development and depict a need for better methodological support on applying UML in legacy development.

In previous works, especially [7,8,10], we describe possible processes and methods to deal with models integrated into a code-centric agile development process, but do not yet provide empirical evidence on its usefulness. Many other works focus only on special issues, e.g. when models are massively used, they need a quality management process. See e.g, references in [1] and [5].

### 3.2 Outline and Approach of the Study

Our objective is to investigate the conditions and challenges for a successful adoption of the model driven development paradigm at Siemens by running a quantitative survey among a high number of software projects. To our knowledge, typical papers and reports on the application of the model driven development paradigm only focus on single projects to gather data, e. g. [17]. By contrast, we want to take a high number of projects into account and assess them regarding to modeling and model driven development. From the results of the survey and analysis of the projects we expect to get hints on the key factors for successful modeling in a large, software developing company. We want to derive possible improvements of the software development processes at Siemens.

For the purpose of preparing the survey we are currently running several guided interviews with project managers from different organizational units of Siemens, located at different sites and concerned with different domains. The outcome of these interviews are used to detect the status quo in modeling and model driven development at Siemens and to derive assumptions and hypotheses that shall be proved in the survey.

### 3.3 Interviews as integral Part of the Study

The qualitative data of the interviews is used to develop the subsequent survey. The interview is set up as a guided interview and covers 54 questions in eight groups:
- Personal data (6 questions): skills, experiences, etc.
- Project information (8 questions): scope, effort, etc.
- Information on the organization and process (5 questions): process model, etc.
- Modeling (14 questions): goals, extent, problems, etc.
- Model Driven Development (7 questions): goals, extent, problems, etc.
- Tools and Technologies (5 questions): usage, problems, etc.
- Team Qualifications (5 questions): education, training, etc.
- Others (4 questions): planning, improvements, etc.

The questions relate to one concrete project chosen by the project manager. The projects should have used the UML as a modeling language and, ideally, used the model driven development paradigm.

In Mid March 2009, six interviews have already been conducted. The interviewees were project managers of (embedded) software projects in the domains of industry and railway automation, automotive and enterprise software. They were guided by the interviewer through the set of open questions, a single interview took up to two hours.

### 3.4 A first Set of Hypotheses

Although there have only been conducted six interviews so far, the qualitative setup of the interviews allows to postulate a first set of hypotheses that shall be validated in the subsequent quantitative survey:
- Models are hardly used for communication or documentation purposes but mostly for generating purposes. Most teams did not model before they introduced model driven development. The introduction of modeling activities is regarded as a necessity of the introduction of the model driven development paradigm. The terms *modeling* and *generating* are often seen as synonyms as the teams use models only for generating artifacts.

- The model driven development paradigm is often used informal, i. e. there are no formal (organizational) processes or methods related to modeling or the model driven development paradigm. The same applies for the modeling tasks. As mentioned above, modeling is regarded just as a supporting activity but not as a full-scale phase with task, role and outcome definitions.
- Projects that successfully adopt the model driven development paradigm are small and agile. The paradigm is not predetermined from the organization or driven by economical considerations but is started from within the teams ("grass roots movement"). There is often one key team member pushing the adoption of the paradigm.
- The most successful adoption of modeling and the model driven development paradigm is achieved when the team members have formal qualifications (e. g. a computer science degree) and are systematically trained in (UML) modeling.
- The UML is often regarded as a too powerful and complex language. Teams that are insufficiently trained in UML modeling, assume that for successful modeling almost all UML elements have to be used.
- Typical and important reasons for adopting the model driven development paradigm are raising the software quality and enforcement of consistent structures and architectures.

Besides these hypotheses we also want to consider some of the results and hypotheses from the related work (especially from [16,17]).

### 3.5 Survey

Based on the insights we gained from the interviews we are developing a quantitative survey to validate respectively falsify the hypotheses we postulated as well as possibly further hypotheses from related work. As most questions in the interview are open and hardly to quantify, we have to adopt and transform them to multiple choice questions.

Target group of the survey are software projects which used (UML) modeling or even used model driven development techniques. The problem is that we will not cover conventional projects this way, but we are interested in targeting successful as well as challenged ones, small and large ones and a variety of domains. We want to assert the constraints for successful model driven projects and identify process improvements from a foundational perspective. We want to address project managers from all three sectors of Siemens and estimate a return of at least 100 results which hopefully will be good enough for quantitative conclusions.

We are aware that a pre-selection of the target group could be a threat to the survey, as well as running the survey in a single company. But we assume that Siemens is big enough to be representative for at least other big software developing companies.

## 4 Conclusion and Outlook

We do know, that model driven development today by far does not yet deliver its promises. Various variants of the paradigm, assisted by various, but often in their functionality very similar tools are not that easy and effective to use, such that quality increases, time-to-market and costs at the same time can be reduced. A number of papers have already discussed that, but we still do not profoundly know what the actual obstacles are. Therefore, we have started this effort and reported on its current status. We will hope that the results help us understanding how to really improve model driven development and will have impact to larger parts of Siemens as well as possibly other German and European industrial companies with heavy parts of software.

## Acknowledgements

We are thankful to Prof. Lutz Prechelt from FU Berlin for discussions and critically reviewing our approach from an empirical perspective.

## References

1. Fieber, F., Huhn, M., Rumpe, B.: Modellqualität als Indikator für Softwarequalität: eine Taxonomie. Informatik Spektrum /2008, Springer, Berlin (2008)
2. France, R., Rumpe, B.: Model-Driven Development of Complex Software: A Research Roadmap. In: Future of Software Engineering 2007 at ICSE. Minneapolis, pg. 37-54, IEEE, May 2007.
3. Petrasch, R., Fieber, F.: Model-Driven Architectur (MDA) in Verbindung mit dem V-Modell XT. In: Petrasch, R., Fieber, F., Macos, D., Bohlen, M. (eds.): Schriften zum Software-Qualitätsmanagement, Band 4, pp 51-60, Logos, Berlin (2008)
4. Petrasch, R., Fieber, F., Meimberg, O., Morlok, S.: Model-Driven Architectur (MDA) im Kontext von Vorgehensmodellen. In: Cremers, A.B., Manthey, R., Martini, P., Steinhage, V. (eds.): INFORMATIK 2005 - Informatik LIVE!, Band 2 (2005)
5. Frank, U., Rumpe, B.: Qualität konzeptueller Modelle. Workshop Modellierung 2006, 21.–24. März 2006, Innsbruck (2006)
7. Grönniger, H., Krahn, H., Rumpe, B., Schindler, M.: Integration von Modellen in einen codebasierten Softwareentwicklungsprozess. In: Mayr HC, Breu R (Hrsg) Modellierung 2006, 22.–24. März 2006, Innsbruck. Lecture Notes in Informatics (LNI), GI-Edition (2006)
8. Rumpe, B.: Agile Modellierung mit UML – Codegenerierung, Testfälle, Refactoring. Springer, Heidelberg (2004)
9. Rumpe, B. :Modellierung mit UML – Sprache, Konzepte und Methodik. Springer, Heidelberg (2004)
10. Rumpe, B.: Agile modeling with the UML. In: Wirsing, M., Knapp, A., Balsamo, S. (eds.): Radical innovations of software and systems engineering in the future. 9th International Workshop, RISSEF 2002, 7.–11. Oktober 2002, Venice, Italy. LNCS 2941 (2004)
11. Margull, U., Kersten, M. and Regnat, N.: A Model-Based Development Method for Device Drivers. In: Hegering, H.-G., Lehmann, A., Ohlbach, H.J. and Scheideler, C. (eds.): INFORMATIK 2008, Beherrschbare Systeme - dank Informatik, Band 2, Beiträge der 38.

Jahrestagung der Gesellschaft für Informatik e.V. (GI), 8. - 13. September, in München, pp. 656-661, GI (2008)
12. Lange, C.: Assessing and Improving the Quality of Modeling - A Series of Empirical Studies about the UML. Dissertation, Technische Universiteit Eindhoven, 2007, IPA dissertation series 2007-14 (2007)
13. Wang, W., Brooks, R.J.: Empirical Investigations of Conceptual Modeling and the Modeling Process. In: Henderson, S.G., Biller, B.,Hsieh, M.-H.,Shortle, J., Tew, J.D. Barton,R.R. (eds.): Proceedings of the 2007 Winter Simulation Conference. IEEE (2007)
14. Dzidek, W.J., Arisholm, E., Briand, L.C.: A Realistic Empirical Evaluation of the Costs and Benefits of UML in Software Maintenance. IEEE Transactions on Software Engineering, vol. 34, no. 3, pp. 407-432 (2008)
15. Anda, B., Hansen, K.: A Case Study on the Application of UML in Legacy Development. ROA'06 May 21, 2006, Shanghai (2006)
16. Asadi, M., Ramsin, R.: MDA-Based Methodologies: An Analytical Survey. In: I. Schieferdecker and A. Hartman (Eds.): ECMDA-FA 2008, LNCS 5095, pp. 419-431 (2008)
17. Mohagheghi, P., Dehlen, V.: Where Is the Proof? - A Review of Experiences from Applying MDE in Industry. In: I. Schieferdecker and A. Hartman (Eds.): ECMDA-FA 2008, LNCS 5095,, pp. 432-443 (2008)
18. OMG: MDA Guide Version 1.0.1. http://www.omg.org/docs/omg/03-06-01.pdf, May 9th 2009
19. Petrasch, R., Meimberg, O.: Model Driven Architecture. dpunkt.verlag, Heidelberg (2006)
20. OMG: UML Infrastructure. http://www.omg.org/spec/UML/2.2/, May 9th 2009
21. OMG: UML Superstructure. http://www.omg.org/spec/UML/2.2/, May 9th 2009